\documentclass[twocolumn,amsmath,amssymb,floatfix,prb,showpacs,footinbib]{revtex4}
\usepackage{amsmath,amssymb,natbib,bm,graphicx,url,psfrag,epsfig,times}
\usepackage[ansinew]{inputenc}

\newcommand{\be}{\begin{equation}}
\newcommand{\ee}{\end{equation}}
\newcommand{\bes}{\begin{equation}\begin{split}}
\newcommand{\ees}{\end{split}\end{equation}}

\newcommand{\vc}[1]{\mathbf{#1}}

\newcommand{\abs}[1]{\left|#1\right|}
\newcommand{\bra}[1]{\left\langle \, #1 \,\right|}
\newcommand{\ket}[1]{\left|\, #1 \, \right\rangle}

\begin{document}
\title{Effects of charge-dependent vibrational frequencies and anharmonicities in transport through molecules}
%Broadening of phonon steps due to charge-dependent vibrational frequencies and anharmonicities in single-molecule devices}
\author{Jens Koch}
\author{Felix \surname{von Oppen}}
\affiliation{Institut f\"ur Theoretische Physik, Freie Universit\"at Berlin, Arnimallee 14, 14195 Berlin, Germany}
\date{July 25, 2005}
\begin{abstract}
As a step towards a more realistic modeling of vibrations in single-molecule devices, we investigate the effects of charge-dependent vibrational frequencies and anharmonic potentials on electronic transport. For weak phonon relaxation, we find that in both cases  vibrational steps split into a multitude of substeps. This effectively leads to a bias-dependent broadening of vibrational features in current-voltage and conductance characteristics, which provides a fingerprint of nonequilibrium vibrations whenever other broadening mechanisms are secondary. In the case of an asymmetric molecule-lead coupling, we observe that frequency differences can also cause negative differential conductance. 
\end{abstract}
\maketitle
\emph{Introduction.}---Recently, several experiments have probed the electron-phonon coupling in single-molecule devices by detecting vibrational features in current-voltage characteristics ($IV$s).\cite{park,ruitenbeek,ho,yu} This coupling and its consequences for electronic transport through molecules have also been at the focus of theoretical studies of $IV$s\cite{schoeller3,flensb1,flensb2,varma,nitzan2,aleiner} and noise spectra,\cite{aleiner,koch2} and it is perceived as a possible avenue towards the design of molecular devices.\cite{domcke}

The signatures of electron-phonon coupling differ depending on the transport regimes. Vibrational steps appear in $IV$s in the sequential-tunneling regime,\cite{schoeller3,flensb1,aleiner,koch2} and in the conductance $dI/dV$ in the inelastic-cotunneling regime.\cite{varma} The visibility of these steps generally depends on the step heights and spacings, as well as their broadening. Three main mechanisms for broadening of vibrational steps in $IV$ and conductance of single-molecule devices have been identified: Broadening induced by (i) temperature,  (ii) vibrational dissipation,\cite{flensb1} and (iii) molecule-lead tunneling.\cite{flensb2} In this paper, we discuss an additional broadening mechanism which arises when going beyond the one-mode harmonic approximation of previous models.

Essentially, all studies to date, e.g. Refs.~\onlinecite{schoeller3,flensb1,flensb2,varma,aleiner,koch2}, are based on two simplifying approximations. First, they restrict the description of the molecular vibrations to the harmonic approximation, and second, they  assume the vibrational frequencies to be identical for the different molecular charge states relevant for the transport. In devices with symmetric voltage splitting, the combination of these two approximations leads to \emph{strictly equidistant} steps in $IV$ for specific gate voltages. 

Here, we take a step towards a more realistic modeling of the vibrations by investigating both the case of  vibrational frequencies depending on the molecular charge state, and the case of anharmonic oscillations within a Morse-potential model. Using a rate-equations approach valid for the regime of weak molecule-lead coupling, we calculate $IV$s and conductances, and show that strictly equidistant vibrational steps are indeed an artefact of the simplifications in previous models. Remarkably, we find that the strength of direct vibrational relaxation can significantly alter the current-voltage characteristics. For strong relaxation, the extended models mainly lead to shifts of the step positions. In contrast, for weak vibrational relaxation they effectively result in a \emph{bias-dependent broadening} of vibrational steps due to the steps splitting  into a multitude of closely spaced substeps.

\emph{Model.}---Our starting point is a generic model for a molecule coupled to metallic leads.\cite{glazman2,schoeller3,aleiner,koch} Transport is assumed to be dominated by one spin-degenerate electronic level with energy $\varepsilon$ (measured with respect to the zero-bias Fermi energies of the leads), tunable by a gate electrode. The system is described by the Hamiltonian $H= H_\text{mol} + H_\text{leads} + H_\text{mix}$, where $H_\text{leads}= \sum_{a=L,R}\sum_{\vc{p},\sigma} \epsilon_\vc{p}c^\dag_{a\vc{p}\sigma}c_{a\vc{p}\sigma}$ describes the non-interacting leads, $H_\text{mix}= \sum_{a=L,R}\sum_{\vc{p},\, \sigma} \left( t_a c^\dag_{a\vc{p}\sigma} d_\sigma + \text{h.c.}\right)$ the tunneling between leads and molecule, and 
\begin{align}
H_\text{mol}= &\varepsilon n_d + \frac{U}{2} n_d(n_d-1)+ \frac{P^2}{2\mu} + V_{n_d}(X)  \label{Hmol}
\end{align}
models the molecular degrees of freedom. Here, $U$ is the charging energy for double occupation,  $d_\sigma$ ($d_\sigma^\dag$) annihilates (creates) an electron with spin projection $\sigma$ on the molecule, and $n_d=\sum_\sigma d_\sigma^\dag d_\sigma$ denotes the corresponding occupation-number operator. Similarly, $c_{a\vc{p}\sigma}$ ($c_{a\vc{p}\sigma}^\dag$) annihilates (creates) an electron in lead $a$ ($a=L,R$) with momentum $\vc{p}$ and spin projection $\sigma$.

The kinetic and potential energy terms in Eq.~\eqref{Hmol} refer to the nuclear motion.  For definiteness, we consider a model with  one dominant mode of vibrations. Then, $X$, $P$, and $\mu$ denote the corresponding normal coordinate, momentum, and reduced mass, respectively. For the singly and doubly charged molecular ions, the potential energy curves generally deviate from their neutral counterpart. In the spirit of the Born-Oppenheimer approximation, we take this into account by writing the potential energy in the form $V_{n_d}(X) = \sum_n v_n(X) \ket{n}\bra{n}$, using projectors onto the electronic ground states with fixed molecular charge $n=0,1,2$.

The coupling between molecule and leads, parameterized by the tunneling matrix elements $t_L$ and $t_R$, is assumed to be weak, i.e.~the tunneling-induced energy broadening $\gamma$ of electronic levels is small, $\gamma\ll k_BT,\hbar\omega_0$, where $T$ denotes the temperature, and $\omega_0$ the vibrational frequency. Then,  $H_\text{mix}$ can be treated perturbatively and the solution of rate equations is sufficient for current and noise calculations.\cite{flensb1,varma,aleiner} For simplicity, we assume symmetric voltage splitting throughout this paper, and focus on the case of strong Coulomb blockade ($U\rightarrow\infty$).

In the absence of tunneling, the molecular states can be written as $\ket{n,q}$ with $n$ denoting the charge state of the molecule and $q$ the number of excited phonons. The rates $W^{n\rightarrow n'}_{q\rightarrow q'}=\sum_{a=L,R}W^{n\rightarrow n'}_{q\rightarrow q';\,a}$ for transitions $\ket{n,q}\rightarrow\ket{n',q'}$ are calculated via Fermi's golden rule. The occupation probabilities $P^n_q$ are determined by the rate equations
\begin{align}\label{rateeq} 
\frac{dP^n_q}{dt}= &\sum_{n',q'}  \left[ P^{n'}_{q'} W^{n'\rightarrow n}_{q'\rightarrow q} - P^{n}_{q} W^{n\rightarrow n'}_{q\rightarrow q'} \right]\nonumber\\
&-\frac{1}{\tau}[{\textstyle P^n_q- P^\text{eq}_q \sum_{q'} P^n_{q'}}],
\end{align}
where the last term describes direct vibrational relaxation towards the equilibrium phonon distribution $P^\text{eq}_q=[1-e^{-\beta\hbar\omega_0}]e^{-\beta q\hbar\omega_0}$ with rate $1/\tau$. The steady-state current is calculated by solving the rate equations \eqref{rateeq} in the stationary case and evaluating
\be\label{current}
I=e\sum_{n,q,q'}P^{n}_{q} \left[W^{n\rightarrow n-1}_{q\rightarrow q';\,R} - W^{n\rightarrow n+1}_{q\rightarrow q';\,R} \right]. 
\ee

Recent papers investigating the influence of molecular vibrations on current and noise have focused on the model  $v_n(X)=\frac{1}{2}\mu\omega_0^2 (X+\sqrt{2}n\lambda\ell_\text{osc})^2$, i.e.~vibrations are taken into account within the harmonic approximation and their frequencies are assumed to be independent of the molecular charge state.\cite{schoeller3,flensb1,flensb2,varma,aleiner,koch2} The position of the potential minimum explicitly depends on the charge state, and its shift, measured in units of the oscillator length $\ell_\text{osc}=(\hbar/\mu\omega_0)^{1/2}$, is characterized by the electron-phonon coupling strength $\lambda$.

We extend this model as follows. In the first case \eqref{one}, we include frequency variations for the different charge states by using 
\be\label{one}\tag{M1}
v_n(X)=\frac{1}{2}\mu\omega_n^2 (X+\sqrt{2}n\lambda\ell_\text{osc})^2.
\ee
 In the second case \eqref{two}, we investigate the effects of anharmonicities within a model based on the Morse potential\cite{morse}
\be\label{two}\tag{M2}
v_n(X)=D\left\{ (1-e^{-\beta(X-\sqrt{2}n\lambda\ell_\text{osc})})^2-1\right\},
\ee
where $D$ denotes the dissociation energy, $\beta$ determines the inverse range of the potential, and $\ell_\text{osc}=(\hbar/\mu\omega_e)^{1/2}$ characterizes the range of the ground-state wave function with $\omega_e=\beta(2D/\mu)^{1/2}$. The number of bound states is given by $\lceil j \rceil+1$ where $j=\sqrt{2\mu D}/\beta\hbar-1/2$.
The Morse potential encompasses both bound and continuum states and is therefore also suitable to describe dissociation processes.\cite{koch4}  In this paper, we restrict the discussion to 
%sufficiently deep potentials, low bias voltages, and intermediate electron-phonon coupling so that 
situations in which transitions into continuum states can be neglected. 

An important qualitative difference compared to the harmonic approximation arises from the \emph{asymmetry} of the Morse potential under parity transformations $X\rightarrow-X$, which causes the overlap of vibrational wavefunctions and hence the transition rates to depend on the \emph{direction} of the shift of the potential minima. In contrast to the case of symmetric potentials, the FC matrix elements now behave differently depending on the sign of the electron-phonon coupling $\lambda$.\cite{iachello} (For symmetric potentials the sign is irrelevant and $\lambda$ can be chosen to be a positive number.) In the general case, this leads to a dependence of the stationary current on the sign of $\lambda$.

\emph{Calculations.}---The calculation of the transition rates $W^{n\rightarrow n'}_{q\rightarrow q'}$ essentially reduces to the evaluation of Franck-Condon (FC) matrix elements,\cite{koch} given by the overlap of the oscillator wave functions $\phi_{n,q}$ of the initial and final states $\ket{n,q}$ and $\ket{n',q'}$,
\be
M^{n\rightarrow n'}_{q\rightarrow q'}=\int_{-\infty}^\infty dx\, \phi_{n,q}^*(x)\phi_{n',q'}(x).
\ee
In harmonic models with $\omega_0=\omega_1=\omega_2$ the FC matrix elements are independent of $n,n'$ and can be compactly written in terms of Laguerre polynomials, see e.g.~Ref.~\onlinecite{koch}. For the harmonic model \eqref{one} with different vibrational frequencies $\omega_n$, and the Morse potential model \eqref{two}, no such simple expression is available\footnote{We point out that an analytical solution of the integral in Eq.~\eqref{morseint} in terms of an alternating sum is possible.\cite{iachello} However, for potentials with a large number of bound states the sum is intractable, and direct numerical quadrature is preferrable.} and instead we evaluate the integrals
\begin{widetext}
\begin{align}\label{omint}
M^{n\rightarrow n'}_{q\rightarrow q'}&=\frac{(\alpha_{n}\alpha_{n'})^{1/4}}{(2^{q+q'}q!q'!\pi)^{1/2}}
\int_{-\infty}^\infty d\xi\, H_q(\alpha_n\xi+\alpha_nn\sqrt{2}\lambda)H_{q'}(\alpha_{n'}\xi+\alpha_{n'}n'\sqrt{2}\lambda)
e^{-\frac{1}{2}(\alpha_n\xi+\alpha_nn\sqrt{2}\lambda)^2-\frac{1}{2}(\alpha_{n'}\xi+\alpha_{n'}n'\sqrt{2}\lambda)^2},\\
M^{n\rightarrow n'}_{q\rightarrow q'}&=2\left[\frac{q!q'!(j-q)(j-q')}{\Gamma(2j-q+1)\Gamma(2j-q'+1)}\right]^{1/2}a^{q'-q}\int_0^\infty d\xi\,
\xi^{2j-q-q'-1}L_q^{2(j-q)}(a\xi)L_{q'}^{2(j-q')}(a^{-1}\xi)e^{-\frac{a+a^{-1}}{2}\xi}\label{morseint}
\end{align}
\end{widetext} 
numerically, where Eqs.~\eqref{omint} and \eqref{morseint} correspond to \eqref{one} and \eqref{two}, respectively. $H_q(x)$ denotes the Hermite polynomial of order $q$, $L^\alpha_n(x)$ the generalized Laguerre polynomial,  $\alpha_n=\sqrt{\omega_n/\omega_0}$ and $a=e^{-\frac{1}{2}\beta(n'-n)\sqrt{2}\lambda\ell_\text{osc}}$. 

Since the transition rates $W^{n\rightarrow n'}_{q\rightarrow q'}$ are bounded from above, and the stationary occupation probabilities obey $P^n_q\to0$ for $q\to\infty$, the Eqs.~\eqref{rateeq} and \eqref{current} can be effectively  solved for a finite number of relevant phonon excitations. For the Morse potential, this treatment requires the restriction to a sufficiently high number of bound states and sufficiently low voltages so that all relevant excitations remain far below the dissociation threshold, $q\ll\lceil j \rceil$.

\begin{figure}
	\centering
 \includegraphics[width=1.0\columnwidth]{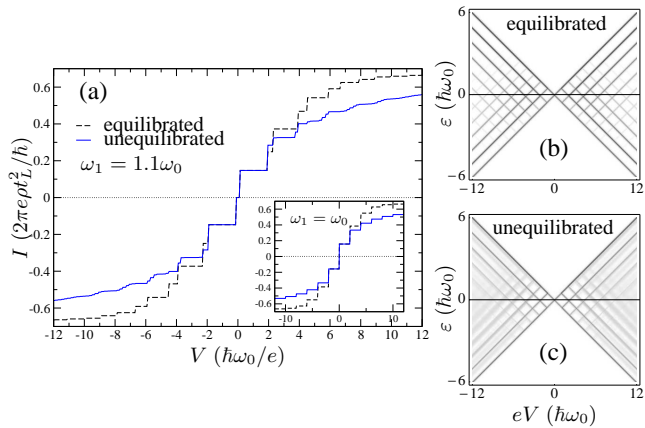}
	\caption{(Color online) $IV$ and conductance for a symmetric device ($t_L=t_R$, symmetric voltage splitting) with slightly different vibrational frequencies for the neutral and charged molecule ($\omega_1=1.1\omega_0$), and $\lambda=1.2$, $U\rightarrow\infty$. 
	(a) $IV$ at $\varepsilon=0$ and $k_BT\ll\hbar\omega_0$ for strong and weak vibrational relaxation.
	While for equilibrated phonons, step locations are specified by Eq.~\eqref{pos1},
	for unequilibrated phonons a multitude of steps is generated, leading to an effective step-broadening even at low temperatures. For comparison, the inset shows results with identical frequencies $\omega_0=\omega_1$. 
	(b),(c) Conductance plots at $k_BT=0.02\hbar\omega_0$ showing the subsplitting and washing out of phonon features for unequilibrated phonons.
 \label{fig1}}	
\end{figure}

\emph{Results and interpretation.}---We first discuss results for model \eqref{one} with harmonic vibrations and charge-dependent frequencies. Representative results are depicted in Fig.~\ref{fig1}. For equilibrated phonons ($\tau\to0$) and frequencies $\omega_0\not=\omega_1$, steps appear at different positions than in the identical-frequencies case, see Fig.~1(a). In principle, steps remain well-defined as shown in the corresponding conductance plot, see Fig.~\ref{fig1}(b), where only small shifts of the series of vibrational sidebands are caused by the entering of two different energy scales $\omega_0$ and $\omega_1$. In contrast, for unequilibrated phonons ($\tau\to\infty$) a splitting of vibrational steps into a multitude of substeps is observed to cause an effective broadening, which increases with bias voltage, see Fig.~\ref{fig1}(a),(c). We now give a more quantitative explanation for this behavior.

%Vibrational structures in $IV$s and conductance plots reflect changes in the transition rates as a function of gate and bias voltage. This dependence is purely due to the leads' Fermi factors in the golden-rule rates, and at low temperatures gives rise to steplike increases of the current whenever new excitation processes become energetically permissible.

For equilibrated phonons and temperatures $\gamma \ll k_BT\ll\hbar\omega_{0,1}$, the phonon distribution $P_q=\sum_n P^n_q$ is strongly peaked for the phonon ground state $q=0$. Consequently, phonon transitions predominantly occur in the channel $0\to q$. Energy conservation limits the possible excitation processes, and in the $(eV,\epsilon)$-plane the boundaries at which the transitions $\ket{0,0}\to\ket{1,q}$ and $\ket{1,0}\to\ket{0,q}$ become possible are given by
\be\label{pos1}
\varepsilon = \mp eV/2 + E^0_0 - E^1_q,\quad  \text{and} \quad 
\varepsilon = \mp eV/2 + E^0_q - E^1_0,
\ee
where $E^n_q=\hbar\omega_n(q+1/2)$ denotes the phonon energy of the state $\ket{n,q}$, and the lower (upper) sign refers to the left (right) lead. In addition, current only flows if electrons can traverse the molecule, leading to the condition 
\be\label{cond}
\abs{eV/2}\ge \abs{\varepsilon-E^0_0+E^1_0},
\ee
which together with Eq.~\eqref{pos1} completely specifies the positions of all vibrational sidebands for equilibrated phonons. In particular, even small frequency differences result in $IV$s with non-equidistant phonon step spacings,\footnote{We remark that model \eqref{one} can still yield $IV$s with equidistant phonon steps for equilibrated phonons, if \emph{asymmetric} voltage splitting (not considered here) is introduced.} and the central crossing point is now located at $(eV,\varepsilon)=(0,\hbar[\omega_0-\omega_1]/2)$ due to the deviation of the zero-point energies of the oscillators.

In contrast, for unequilibrated phonons  a multitude of excitation and de-excitation processes $\ket{0,q}\to\ket{1,q'}$ and $\ket{1,q}\to\ket{0,q'}$ are permitted, and Eq.~\eqref{pos1} must be replaced by
\be\label{pos2}
\varepsilon = \mp eV/2 + E^0_q - E^1_{q'},\quad  \text{and} \quad 
\varepsilon = \mp eV/2 + E^0_{q'} - E^1_q.
\ee
However, the range of the contributing $q,q'$ in Eq.~\eqref{pos2} now inherently depends on the nonequilibrium distribution $P^n_q$, which makes concise statements about the occurence of certain steps more difficult. Generally, larger bias voltages $\abs{eV}$ cause the nonequilibrium distribution to widen, and thus increase the range of relevant $q,q'$ in Eq.~\eqref{pos2}. This corresponds to a splitting of a vibrational step into an increasing number of closely spaced substeps as observed in Fig.~\ref{fig1}(a),(c), resulting in an effective broadening of steps.

\begin{figure}
	\centering
		\includegraphics[width=0.6\columnwidth]{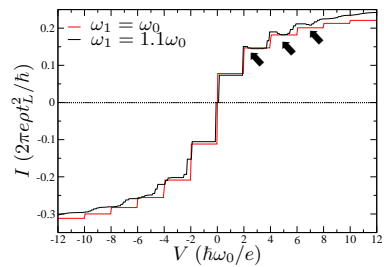}
	\caption{(Color online) Current-voltage characteristics for an asymmetric device ($t_R^2=0.3t_L^2$, symmetric voltage splitting)  in the unequilibrated regime with $\lambda=1.1$, $k_BT\ll\hbar\omega$, $U\rightarrow\infty$. The combination of asymmetric coupling and differing vibrational frequencies is observed to lead to weak peaklike structures at the onsets of phonon steps with negative differential conductance regions (marked by arrows).
 \label{fig2}}	
\end{figure}

Interestingly, for asymmetric devices ($t_L\not=t_R$) we find that frequency differences between molecular charge states can induce negative differential conductance (NDC).  A representative example for this behavior is shown in Fig.~\ref{fig2}. The splitting of phonon steps causes  peaklike structures at the onset of phonon steps. For certain parameters, such NDC features have been predicted for asymmetric coupling and harmonic vibrations with $\omega_0=\omega_1$ in Refs.~\onlinecite{schoeller3}. Here, we report that qualitatively similar NDC behavior can arise from differing vibrational frequencies, for parameters where our numerical results do not exhibit NDC for $\omega_0=\omega_1$. It is interesting to note that NDC at onsets of phonon steps is also observed in the experiment by Park \emph{et al.}\cite{park,mceuen}

\begin{figure}
	\centering
    \includegraphics[width=1.0\columnwidth]{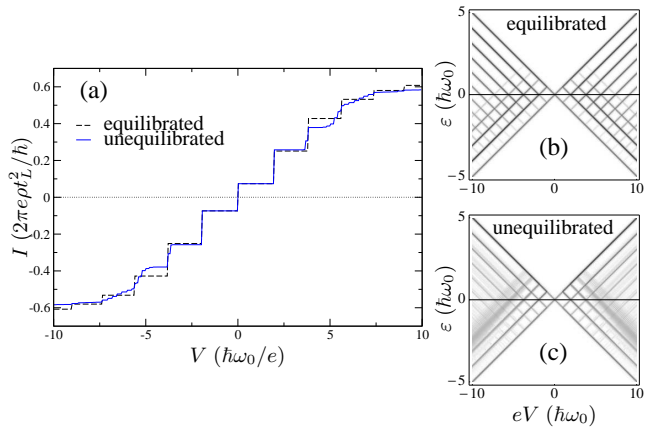}
	\caption{(Color online) $IV$ and conductance plots for a symmetric device ($t_L=t_R$, symmetric voltage splitting) with a Morse potential ($j=30$), for $\lambda=1.5$, $U\rightarrow\infty$. 
	(a) $IV$ at $\varepsilon=0$ and $k_B T\ll\hbar\omega_e$ for strong and weak vibrational relaxation.
	For equilibrated phonons steps along $\varepsilon=0$ are located at voltages specified by Eq.~\eqref{pos1},
	for unequilibrated phonons the step-splitting leads to an effective step-broadening even at low temperatures. 
	(b),(c) Conductance plots at $k_BT=0.02\hbar\omega_e$ showing the subsplitting and broadening of phonon features for unequilibrated phonons.
 \label{fig3}}	
\end{figure}

Results for the model \eqref{two} with anharmonic vibrations are depicted in Fig.~\ref{fig3}. While the main effects are similar to those of model \eqref{one} -- steps in $IV$ being non-equidistant, weak vibrational relaxation leading to the splitting of steps into  substeps -- the underlying mechanism is different. In the Morse model \eqref{two}, the potential curves are shifted depending on the molecular charge state, but are identical otherwise. Here, the appearance of different step spacings and splitting of steps is purely due to the fact that the energy spectrum of an anharmonic oscillator is not equidistant. 

In analogy to model \eqref{one}, for equilibrated phonons the positions of vibrational sidebands are fixed by substituting the eigenenergies of the Morse potential $E_q=-D+\hbar\omega_e(q+1/2)-\hbar\omega_e\chi(q+1/2)^2$ into Eqs.~\eqref{pos1} and \eqref{cond}. Here, we have used $\chi=(2j+1)^{-1}$. This leads to a conductance plot consisting of two sets of parallel lines with decreasing line distances, reflecting the decreasing spacing of eigenenergies of the Morse potential as the quantum number $q$ increases, see Fig.~\ref{fig3}(a),(b). Since the phonon ground state energies are identical, $E^0_0=E^1_0$, the central crossing point is located at $(eV,\varepsilon)=(0,0)$. 

In the unequilibrated case, the anharmonicity of the potential and the occurence of various phonon excitation and de-excitation processes give rise to a splitting of steps. As before, the range of relevant $q,q'$ in Eq.~\eqref{pos2} depends on bias and gate voltage, leading to a growing number of substeps with increasing voltage, and hence an effective washing out of phonon steps, see Fig.~\ref{fig3}(a),(c). 

\emph{Conclusions.}---As a step towards a more realistic modeling of molecular vibrations and their consequences for electronic transport in single-molecule devices, we have investigated the effects of charge dependences of vibrational frequencies and anharmonicities in the sequential-tunneling regime. Even for small frequency differences and anharmonicities, we find that vibrational step spacings cease to be equidistant. In combination with weak vibrational relaxation, both frequency differences and anharmonicities are shown to lead to a bias-dependent splitting of levels, effectively resulting in a broadening of phonon steps. For asymmetric molecule-lead coupling, we find that this mechanism can also lead to negative differential conductance behavior. We conclude that spacings of vibrational features in $dI/dV$ provide information about charge dependence of vibrational frequencies and anharmonicity of the potential. Whenever other broadening mechanisms play a secondary role, the bias-dependent subsplitting or broadening acts as a fingerprint of nonequilibrium vibrations.

It is interesting to note that the appearance of clusters of vibrational substeps due to the emergence of differing energy scales of phonon excitations resembles the splitting of tunneling peaks observed in transport through ultrasmall metallic grains. In that context, Agam \emph{et al.}\cite{agam} argued that the Coulomb interaction effectively results in a bias-dependent splitting of resonance peaks in the nonequilibrium case, i.e.~for sufficiently slow electronic relaxation.
Finally, we remark that similar arguments also suggest a splitting of the conductance fine structure for magnetic single-molecule devices,\cite{elste} whenever the exchange coupling varies with the molecular charge state.

We thank C. Timm for valuable discussions. This work was supported by the Junge Akademie, Sfb 658, and Studienstiftung des deutschen~Volkes.

\end{document}